\documentclass[10pt,conference]{IEEEtran}
\IEEEoverridecommandlockouts

\usepackage[utf8]{inputenc}
\usepackage[T1]{fontenc}
\usepackage{textcomp}

\usepackage{cite}
\usepackage{amsmath,amssymb,amsfonts}
\usepackage{algorithmic}
\usepackage{graphicx}
\usepackage{xcolor}
\usepackage{url}

\usepackage{listings}
\usepackage{caption}
\usepackage{float}
\usepackage{xcolor}
\usepackage{url}

\newfloat{listing}{htbp}{lop}
\floatname{listing}{Listing}
\DeclareCaptionType{lstfloat}[Listing][List of Listings]
\lstdefinelanguage{Gherkin}{
	morekeywords = {
		Given,
		When,
		Then,
		And,
		Scenario,
		Feature,
		But,
		Background,
		Scenario Outline,
		Examples
	},
  sensitive=true,
  morecomment=[l]{\#},
  numbersep=5pt,
  xleftmargin=2em,
  framexleftmargin=1.5em,
}

\begin{document}

\title{Streamlining Acceptance Test Generation for Mobile Applications Through Large Language Models: An Industrial Case Study\\
}

\author{\IEEEauthorblockN{Pedro Luís Fonseca}
\IEEEauthorblockA{\textit{
Critical TechWorks} and \\
\textit{Faculty of Engineering,} \\ \textit{University of Porto}\\
Porto, Portugal \\
up202008307@up.pt}
\and
\IEEEauthorblockN{Bruno Lima}
\IEEEauthorblockA{\textit{LIACC},
\textit{Faculty of Engineering,} \\ \textit{University of Porto}\\
Porto, Portugal \\
brunolima@fe.up.pt}
\and
\IEEEauthorblockN{João Pascoal Faria}
\IEEEauthorblockA{\textit{INESC TEC},
\textit{Faculty of Engineering,} \\ \textit{University of Porto}\\
Porto, Portugal \\
jpf@fe.up.pt}
}

\maketitle

\begin{abstract}
Mobile acceptance testing remains a bottleneck in modern software development, particularly for cross-platform mobile development using frameworks like Flutter. While developers increasingly rely on automated testing tools, creating and maintaining acceptance test artifacts still demands significant manual effort. To help tackle this issue, we introduce AToMIC, an automated framework leveraging specialized Large Language Models to generate Gherkin scenarios, Page Objects, and executable UI test scripts directly from requirements (JIRA tickets) and recent code changes. Applied to BMW’s MyBMW app, covering 13 real-world issues in a 170+ screen codebase, AToMIC produced executable test artifacts in under five minutes per feature on standard hardware. The generated artifacts were of high quality: 93.3\% of Gherkin scenarios were syntactically correct upon generation, 78.8\% of PageObjects ran without manual edits, and 100\% of generated UI tests executed successfully. In a survey, all practitioners reported time savings (often a full developer-day per feature) and strong confidence in adopting the approach. These results confirm AToMIC as a scalable, practical solution for streamlining acceptance test creation and maintenance in industrial mobile projects.
\end{abstract}


\section{Introduction}

Acceptance Testing (AT) remains a critical and resource-intensive step in delivering high-quality software, especially as mobile ecosystems grow more complex and release cycles accelerate~\cite{watkins2010}.

 This challenge is heightened in cross-platform environments like Flutter, where a single codebase must perform reliably across a wide range of devices and operating systems~\cite{AbuSalim_2021,DeOliveira202335}. While Flutter has emerged as a leading framework for building native-quality apps from a unified codebase~\cite{osdacpdwff}, its flexibility introduces new testing complexities. Traditional tools often struggle with Flutter’s widget-based architecture, and the diversity of target platforms makes comprehensive test coverage increasingly difficult to achieve~\cite{yu2023vision}.

Meanwhile, Large Language Models (LLMs) are proving increasingly adept at code generation and natural language processing~\cite{10.1145/3695988}, yet real-world industrial applications of LLMs to automate AT for mobile applications are only beginning to emerge.


Despite adoption of robust tools like Appium and Espresso, industrial mobile teams face persistent obstacles in effective acceptance testing~\cite{9401983}. Platform fragmentation remains a fundamental challenge for acceptance testing, as variations in screen sizes, OS versions, and hardware significantly increase the number of required configurations~\cite{yu2023vision}. For Flutter, in particular, the widget-based and asynchronous model poses unique problems that traditional testing workflows rarely address natively, often forcing engineering teams to develop significant custom infrastructure~\cite{osdacpdwff}.

Continuous evolution of mobile applications adds further difficulty, with frequent feature changes routinely necessitating extensive reworking of test scripts and infrastructure. The cost of maintaining and updating Page Objects, test scenarios, and ensuring overall compatibility can quickly absorb considerable engineering time, slowing down release cycles and risking quality compromises~\cite{Pan2022910}. These bottlenecks are especially acute in industrial projects, as observed in BMW’s MyBMW app, where large, distributed teams are required to maintain rigorous acceptance validation at scale across hundreds of screens and features.


To address these pressing, industrial-scale obstacles, we introduce AToMIC (Acceptance Testing for Mobile Intelligent Code), an automated workflow that leverages the strengths of recent LLMs to streamline and, where possible, fully automate the generation and maintenance of acceptance test artifacts for cross-platform mobile applications. Rather than replacing established frameworks, AToMIC is designed to work in tandem with existing tools, targeting the most labor-intensive components of AT, including the transformation of requirements (such as JIRA tickets) directly into traceable Gherkin scenarios, the automatic construction and maintenance of Page Objects mapped cleanly to code, and the synthesis of executable UI test scripts ready for immediate use in CI/CD pipelines.

This solution has been validated in a real-world industrial context through a partnership with BMW on the MyBMW application, a highly complex and feature-rich Flutter codebase. Our research goals are threefold: first, to demonstrate the practical feasibility and limitations of LLM-driven AT automation at this scale; second, to provide comprehensive measurement and practitioner feedback on the efficiency, output quality, and actual cost savings delivered in daily industrial workflows; and third, to extract actionable guidance for organizations that may face similar scalability and maintainability challenges.

The results from this industrial evaluation show that automated artifact generation can be reliably integrated into large, established development environments. By aligning requirements analysis, code change detection, navigation mapping, and test artifact generation within a single automated process, AToMIC has delivered tangible reductions in manual effort, while maintaining traceability, quality, and compatibility with the workflows required in production-scale software engineering.

The remainder of this paper is organized as follows. Section II reviews related work in testing automation, LLM applications, as well as, necessary context. Section III details the AToMIC system architecture and implementation. Section IV presents our empirical evaluation methodology and results. Section V concludes with future research directions.

\section{Background and Related Work}

This section provides essential background on AT for mobile applications and reviews current advancements in LLM-driven test automation. We focus on the fundamentals of AT, challenges specific to Flutter-based applications, and the state-of-the-art in LLM-based test generation for mobile apps.

\subsection{Acceptance Testing and Mobile Automation}

AT aims to validate a system's behavior from the end-user perspective. It complements other testing layers such as unit and integration testing. Unit testing focuses on verifying individual components, while integration testing ensures proper interaction between different modules. Acceptance tests, although fewer in number, provide high-level assurance that the application meets business and user requirements.

Black-box testing forms the foundation of AT, evaluating a system solely based on its external behavior without access to internal implementation details \cite{9401983}. Testers design scenarios based on requirements and user stories, concentrating on inputs and expected outputs. This approach is especially relevant in mobile applications, where user experience and business logic must align closely with customer expectations \cite{10.1049/iet-sen.2018.5445}.

The integration of Gherkin\footnote{\url{https://cucumber.io/docs/gherkin/}} grammar has significantly enhanced AT by improving communication between technical teams and stakeholders. Gherkin uses a structured syntax with keywords such as Given, When, and Then to define acceptance criteria. In the Appendix, a real-world example from the MyBMW application is showcased, demonstrating a Gherkin-based testing scenario.

\begin{table}[htbp]
\caption{Comparison of Testing Frameworks for Flutter Projects}
\label{tab:flutter_frameworks}
\centering
\small
\begin{tabular}{|l|c|c|c|}
\hline
\textbf{Feature} & \textbf{Espresso} & \textbf{Appium} & \textbf{Flutter Driver} \\
\hline
Flutter Compatib.      & Limited & Good & Excellent \\
Gherkin Support    & Yes & Yes & Yes \\
Language           & Java/Kotlin & Multiple & Dart \\
Flutter-Spec. APIs & No & Limited & Yes \\
Test Exec. Speed         & Faster & Slower & Faster \\
Cross-platform     & Android only & Yes & Flutter only \\
Learning Curve     & Steep & Moderate & Easy \\
\hline
\end{tabular}
\end{table}

Flutter-based applications present distinct testing challenges compared to traditional web or native mobile testing. Table~\ref{tab:flutter_frameworks} compares testing frameworks for Flutter projects. While Appium offers cross-platform capabilities, Flutter Driver provides the most seamless integration with Flutter applications.

In summary, AT plays an indispensable role in ensuring that mobile applications deliver robust, user-centered functionality across diverse platforms. While frameworks such as Appium\footnote{\url{https://appium.io/docs/en/latest/}}, Espresso\footnote{\url{https://developer.android.com/training/testing/espresso}}, and Flutter Driver\footnote{\url{https://api.flutter.dev/flutter/flutter_driver/}} offer essential tooling to automate this process, the unique challenges posed by Flutter’s architecture and mobile ecosystem fragmentation necessitate continuous innovation. The integration of structured languages like Gherkin enhances clarity and collaboration, yet manual effort in test case creation and maintenance remains a significant bottleneck.

\subsection{Mobile UI Testing in the MyBMW App}

\begin{figure}[htbp]
\centering
\includegraphics[width=1.0\linewidth]{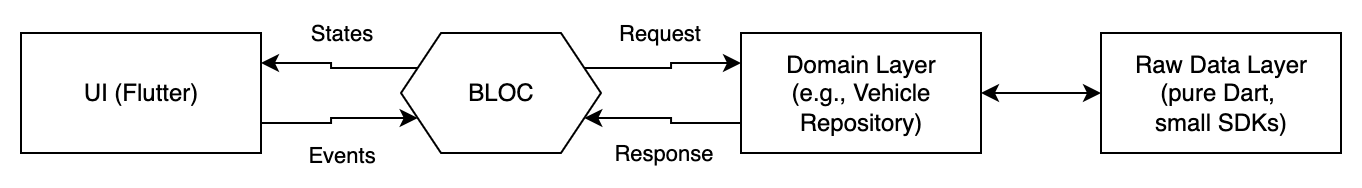}
\caption{MyBMW's Three-Tiered Architecture}
\label{fig:three_tier_architecture}
\end{figure}

The MyBMW app uses a three-tier architecture in Flutter and Dart (see Figure~\ref{fig:three_tier_architecture}). The UI layer, built with Flutter, handles user interaction and presentation, utilizing the BLoC (Business Logic Component) design pattern\footnote{\url{https://pub.dev/packages/bloc}} for state management related to the UI.
The domain layer (pure Dart) contains cross-cutting business logic and coordinates data access via repositories, while the raw data layer connects to external sources like HTTP clients, GPS, and secure storage.
The BLoC pattern separates business logic from the UI, promoting maintainability and a clear separation of concerns~\cite{bmw-arch, bloc-arch}: user events are dispatched to BLoCs, which process them, interact with the domain layer, and emit new states for the UI.

The test automation infrastructure for MyBMW uses Flutter Driver, 
enabling robust interaction with Flutter widgets via unique ValueKey identifiers. UI tests are implemented in Kotlin, following the Page Object design pattern, in which each app screen is abstracted as a dedicated class encapsulating both UI elements and user interactions.

\begin{figure}[htbp]
\centering
\includegraphics[width=1.0\linewidth]{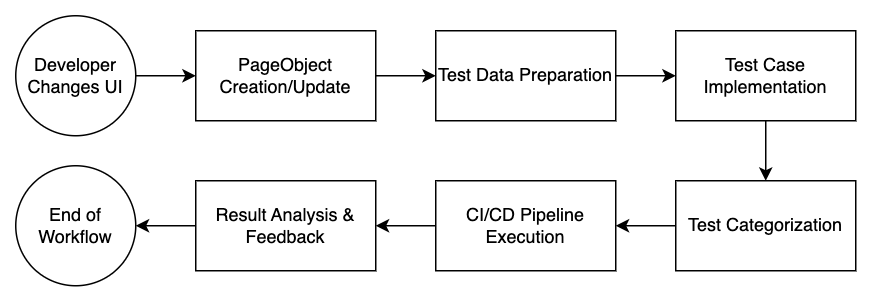}
\caption{UI Test Development Workflow in the MyBMW Project}
\label{fig:ui_test_workflow}
\end{figure}

Figure~\ref{fig:ui_test_workflow} illustrates the standard UI test development workflow used on the MyBMW project. The process starts when developers make UI changes, leading to the creation or update of corresponding Page Object classes that abstract the new or modified screens. Subsequently, relevant test data is prepared, often via comprehensive mocking of backend services, to ensure that tests are reliable and deterministic. Then, concrete test cases are implemented, utilizing the Page Object APIs to describe test scenarios at a high level of abstraction. These test cases are then categorized (e.g., as smoke, regression, or feature tests) and integrated into automated CI/CD pipelines, where they are executed on real or simulated devices. Test results are analyzed and promptly shared with the development team to resolve issues before code reaches production.

Key practices supporting this workflow include assigning unique ValueKeys to widgets for stable selection, using the Page Object pattern to centralize UI logic and reduce redundancy, and employing reliable mock data for consistent test results. Integrating these tests into CI pipelines enables early issue detection and helps maintain release quality.

Despite these best practices, traditional UI test development for Flutter apps remains labor-intensive and error-prone. Keeping Page Objects aligned with changing UI code, preparing test data, and updating scripts as requirements evolve require significant effort. In large projects like MyBMW, this often creates bottlenecks that slow development and limit test coverage, highlighting the need for more efficient, automated acceptance test generation.

\subsection{LLMs in Test Generation for Mobile Apps}

The use of LLMs for automating acceptance test generation has recently gained substantial traction in mobile software development, driven by persistent challenges such as platform fragmentation, rapid release cycles, and increasing application complexity.

XUAT-Copilot represents one of the early industrial-scale applications of LLM-driven automation specifically for mobile user acceptance testing. Deployed in the WeChat Pay app, XUAT-Copilot employs a multi-agent architecture, coordinating agents for state inspection, action planning, and parameter generation. It achieves near-human accuracy, highlighting the practical potential of LLM-based solutions in production environments~\cite{wang2024xuatcopilot}.

VisiDroid extends this domain by combining visual and textual inputs for Android GUI testing. By leveraging multi-modal reasoning, VisiDroid significantly improved task-completion accuracy compared to previous text-only methods, demonstrating the benefits of incorporating visual context into test script generation processes~\cite{huynh2025visidroid}.

LELANTE further emphasizes automated Android testing by employing LLMs to generate and maintain test scripts. Its effectiveness lies in automating test case creation directly from natural language descriptions, offering substantial productivity enhancements and reductions in manual effort. This research underscores the capacity of LLMs to streamline mobile test automation tasks effectively~\cite{fatin2025lelante}.

LLMDroid specifically addresses coverage improvement in mobile GUI testing by guiding automated test case generation through LLMs. It systematically enhances test scenarios, demonstrating measurable improvements in test coverage and scenario effectiveness, reinforcing the potential for LLM guidance in comprehensive mobile GUI testing~\cite{wang2025llmdroid}.


AToMIC stands out from prior mobile testing tools by integrating multiple innovations: unlike XUAT-Copilot and LELANTE, it supports artifact generation with full traceability, derives user flows directly from code structure rather than GUI exploration, and employs specialized LLMs for different subtasks, enabling a novel, privacy-compatible multi-model architecture for industrial CI/CD pipelines.

Distinct from these existing works, AToMIC integrates structured software artifacts such as JIRA issues and GitHub commits into its pipeline. Focusing specifically on Flutter applications, it produces structured and maintainable acceptance test artifacts, including Gherkin scenarios, Page Object classes, and UI test scripts, optimized for seamless integration into industrial CI/CD processes. This explicit alignment with real-world industrial practices, as exemplified in BMW’s MyBMW application, directly addresses critical needs around scalability, maintainability, and traceability.

\section{AToMIC System Design}

\subsection{Architecture Overview}

AToMIC is designed as a modular framework that integrates seamlessly with existing development workflows and tools, enabling adoption without major changes to established practices. Its deployment and validation in the MyBMW app demonstrate its effectiveness in industrial settings.

\begin{figure}[htbp]
    \centering
    \includegraphics[width=1.0\linewidth]{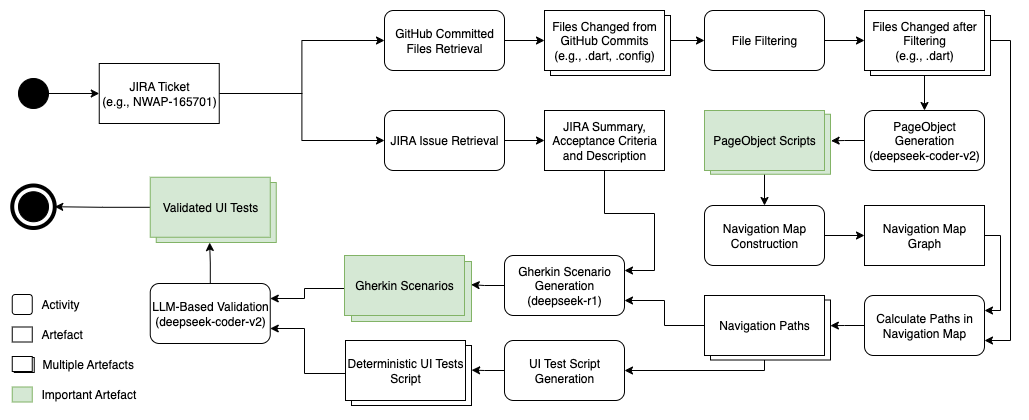}
    \caption{High-Level Activity Diagram of the AToMIC System}
    \label{fig:atomic-activity-diagram}
\end{figure}

Figure~\ref{fig:atomic-activity-diagram} presents the workflow for AToMIC. The workflow begins with JIRA issue specification, where stakeholders define requirements using standard templates and acceptance criteria. The system then retrieves associated code changes from GitHub commits, analyzing modifications to identify affected UI components and navigation flows.

The system then constructs a navigation map that models possible user flows through the application. This graph-based representation captures the relationships between screens and user actions, providing essential context for test generation.

Finally, AToMIC generates three types of artifacts: Gherkin scenarios that formalize acceptance criteria, Page Object classes that abstract UI interactions, and executable test scripts that can be integrated into CI/CD pipelines.

\subsection{LLM Integration Strategy}

Privacy considerations in industrial settings often preclude the use of cloud-based LLM services. Our approach addresses this constraint by utilizing local LLM deployment through Ollama\footnote{\url{https://ollama.com/}}, enabling secure processing of proprietary code and requirements. Within the available open-source options, we conducted internal evaluations of several candidates, selecting a configuration that provided the best balance between accuracy, performance, and efficiency under these constraints.

We employ different models for different tasks, optimizing for the specific requirements of each workflow component. DeepSeek-R1 handles complex reasoning tasks like Gherkin scenario generation~\cite{deepseek-r1}, while DeepSeek-Coder-V2 focuses on code generation for Page Objects and test scripts validation~\cite{deepseek-coder-v2}. Gemma3:1b provides efficient code summarization capabilities~\cite{gemma3-1b}. This specialization approach proved crucial for maintaining reasonable performance with local deployment. By matching model capabilities to task requirements, we achieve better results than using a single general-purpose model for all tasks.

Finally, AToMIC’s modular architecture enables straightforward integration of alternative LLMs, ensuring that performance improvements from newer models can be readily incorporated without requiring architectural changes.

\subsection{Requirements Processing}

The system extracts structured information from JIRA issues, including summaries, acceptance criteria, and labels. This information provides the business context necessary for generating meaningful test scenarios.

Code change analysis locates all modified files and applies project-specific filters to focus on user-facing functionality. Files irrelevant to the UI, such as those in \texttt{/test/}, \texttt{/utils/}, \texttt{/repository/}, or configuration files, are excluded, while Dart files within main feature clusters are prioritized. This filtering approach, based on established directory and naming conventions, ensures that all relevant screen and widget files are captured. 

\subsection{Page Object Generation}

\begin{figure}[htbp]
    \centering
    \includegraphics[width=1.0\linewidth]{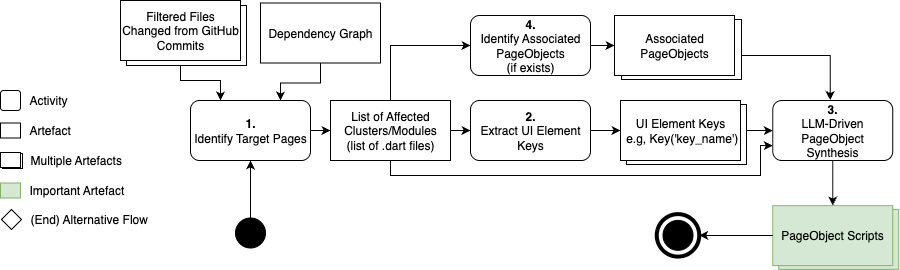}
    \caption{Activity Diagram for Page Objects Generation}
    \label{fig:pageobject-diagram}
\end{figure}


Maintaining alignment between evolving Flutter UI code and test abstractions is a persistent challenge in mobile automation. Our Page Object generation system addresses this by enforcing strict naming and structural conventions, combined with automated analysis and code generation.


Page Objects serve as an abstraction layer between automated tests and the Flutter UI, each representing a screen with methods for user interactions. In large-scale apps like MyBMW, standardized Page Objects are essential for maintainability, traceability, and automated navigation reasoning.



To ensure consistency, we require each Dart UI screen file to end with \texttt{\_page.dart}, and its corresponding Kotlin Page Object to use the same CamelCase name (e.g., \texttt{profile\_page.dart} maps to \texttt{ProfilePage.kt}). Directory structures are mirrored between the app and test code, and widget keys follow a structured format (\texttt{context\_keyIdentifier\_pageItGoesTo}) to embed navigation metadata for automated extraction.

The system features a fully automated Page Object generation and update pipeline. When changes occur in the Flutter codebase, the system identifies which \texttt{\_page.dart} files are affected and, using a precomputed dependency graph built from all Dart files in the MyBMW app, traces the widgets that make up each page—even when they are distributed across multiple files. This dependency graph operates at industrial scale: for the MyBMW application, with approximately 3 million lines of code, it models more than 36,000 Dart files as nodes. 
The graph enables the system to systematically resolve all UI components and dependencies for any page. It then extracts the relevant widget keys used to identify UI elements for testing.

At this point, the system checks whether a Page Object already exists. If it does, it updates the Page Object with the new elements or logic. If not, it initiates the creation of a new Page Object. For generation, the system passes an example Page Object along with the extracted keys and structural data to an LLM (specifically, DeepSeek-Coder-v2), which synthesizes a compliant Page Object class.

This entire flow is illustrated in Figure~\ref{fig:pageobject-diagram},  automating a typically manual and error-prone process, whilst ensuring test code remains consistent with the rapidly evolving Flutter UI.


\subsection{Navigation Map and Navigation Paths Generation}



Traditional mobile testing frameworks struggle with navigation complexity because they lack a comprehensive understanding of application structure. AToMIC addresses this by constructing a complete navigation model that enables systematic exploration of user journeys.

The navigation map formalizes all possible transitions between screens in the application and is essential for ensuring both coverage and effective test automation. It is implemented as a directed multigraph, where the nodes correspond to Page Objects that represent individual screens (such as \texttt{VehicleTabPage}), and the edges represent navigation actions between screens, as inferred from Page Object methods and widget key conventions. 

The construction algorithm leverages the modular code structure and naming standards.
The full map for MyBMW (v5.5.2) has 174 nodes and 232 edges, with parallel edges representing variant flows. 

Based on this navigation map and the Page Objects affected by code changes, the system then identifies valid navigation paths from the app’s entry point (e.g., \texttt{VehicleTabPage}) to the target screens (corresponding to the affected Page Objects), using a depth-limited breadth-first traversal of the navigation map. The discovered paths are prioritized based on their length, corresponding to the most direct user journeys. 
This ensures comprehensive test coverage across different user flows.

The discovered navigation paths provide essential context for Gherkin scenario generation and UI test script synthesis. 
For a concrete example of navigation paths, see the Appendix.


\subsection{Gherkin Scenario Generation}

\begin{figure}[htbp]
\centering
\includegraphics[width=1.0\linewidth]{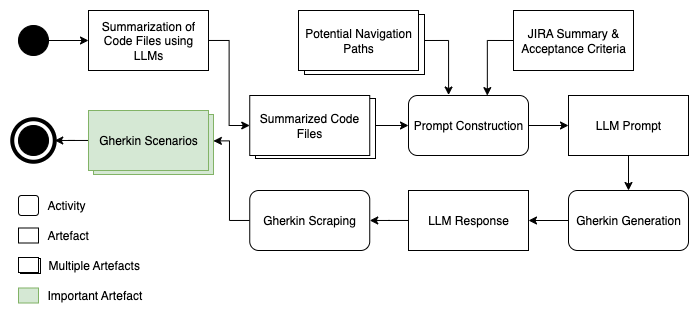}
\caption{Activity Diagram for Gherkin Scenario Generation}
\label{fig:gherkin-generation-diagram}
\end{figure}


Gherkin scenarios act as a crucial bridge between business requirements and technical implementation. Our system automates their generation by combining multiple sources of information in a gray-box approach: JIRA acceptance criteria for business context, code summaries for technical insight, and navigation paths to ensure realistic user flows.

The generation process is LLM-driven and adaptive. It can produce both simple scenarios and more complex edge cases, depending on the depth and richness of the provided inputs. This flexibility is enabled through DeepSeek-R1, which handles natural language synthesis of scenarios based on structured prompts. AToMIC does not apply a hard limit during initial scenario generation. Instead, our approach generates all plausible candidate scenarios and then, at posterior phases in the workflow, applies multi-stage filtering: first, a deduplication process removes semantically overlapping scenarios; second, LLM-based validation eliminates scenarios that are technically infeasible, redundant, or misaligned with the acceptance criteria. This strategy balances comprehensive coverage with practical test suite size, allowing the system to adapt the number of scenarios based on the complexity and scope of each JIRA issue rather than arbitrary limits.

A key supporting component is code summarization, performed using gemma3:1b. Instead of feeding entire source files to the LLM, which can be inefficient and noisy, the system generates concise summaries highlighting relevant functionality and UI elements.

The prompt structure used for LLM generation incorporates several layers of contextual information:

\begin{itemize}
\item \textbf{JIRA fields}, including issue description, acceptance criteria, and labels, form the foundation of business context.
\item \textbf{Code summaries} provide a distilled view of implementation details relevant to the issue.
\item \textbf{Navigation paths}, derived as explained in the previous section, are included to ensure that the scenario aligns with valid user flows in the application.
\end{itemize}

This multi-source prompt enables the LLM to generate Gherkin scenarios that are both technically grounded and aligned with business requirements.

The resulting scenario files serve as formalized acceptance tests. They maintain traceability to JIRA tickets and are directly linked to navigable user paths through the app, facilitating alignment between QA, development, and product stakeholders.

Figure~\ref{fig:gherkin-generation-diagram} presents the full activity flow for this generation process, from input aggregation to final scenario creation.

An example illustrating some of the mentioned artifacts is provided in the Appendix.

\subsection{UI Test Script Generation}


The final stage of the workflow is the generation of executable UI test scripts from Gherkin scenarios and the potential navigation paths. This process is primarily deterministic, translating the identified navigation paths (corresponding to possible user journeys) into Kotlin test functions through a custom-built generator developed for this project. The generator scaffolds necessary imports, manages base class inheritance, and integrates PageObjects to ensure consistency with BMW’s testing framework. A final LLM validation step then refines the results to ensure alignment with the Gherkin scenarios derived from the business requirements (see Figure~\ref{fig:atomic-activity-diagram}).

For each path, the system deterministically generates a Kotlin test function. These functions leverage Page Object methods to traverse the application, access the target screen, and return to the initial state. This approach ensures that each test executes in isolation while accurately reflecting realistic user behavior.

When corresponding Gherkin scenarios are available, a code-specialized LLM (DeepSeek-Coder-V2) performs validation and refinement of the generated test script. It ensures alignment between test steps and the scenarios, removes unrelated logic, and inserts explanatory comments to enhance clarity.

The final test scripts are placed into the appropriate module and are ready for CI/CD execution. 


\section{Empirical Evaluation}
This section presents a comprehensive empirical evaluation of the AToMIC system for automating acceptance test generation in industrial Flutter mobile application development. The study was conducted using BMW's MyBMW app, a production system with over 170 screens.

\subsection{Research Questions}

This evaluation aims to answer four research questions:

\begin{description}
\item[RQ1:] \textbf{Accuracy and Completeness}: To what extent are the generated artifacts accurate and complete when using LLMs for automating AT in mobile apps?

\item[RQ2:] \textbf{Time Efficiency}: Is the process of generating acceptance test scripts using LLMs more time-efficient compared to traditional manual methods?

\item[RQ3:] \textbf{Accessibility}: How accessible and understandable are the generated acceptance test scripts for individuals with varying levels of technical expertise?

\item[RQ4:] \textbf{Challenges and Limitations}: What are the main challenges and limitations encountered when applying LLMs to acceptance test generation for mobile applications, and how do they impact the overall process?
\end{description}


\subsection{Experimental Setup}

The evaluation follows a mixed-methods strategy, combining quantitative analysis with qualitative insights to provide a well-rounded view of AToMIC’s performance.

On the quantitative side, the focus is on measuring artifact quality, generation time, and LLM token usage across real-world tasks. In parallel, qualitative validation was conducted through a survey of MyBMW practitioners, capturing perceptions of AToMIC’s usability, output clarity and usefulness, impact on productivity, trust in the generated tests, and likelihood of integration into regular workflows.

All validation activities were performed using 13 real JIRA issues from the MyBMW Flutter codebase. 
These issues, linked to a total of 67 commits (since organizational policies require every pull request and commit to reference a JIRA issue key), reflect diverse functionality and navigation complexity, ensuring a realistic and representative test set.
Minor adjustments were made to the workflow and code structure by applying the technical standards and naming conventions introduced in this work directly to the files and modules associated with these issues. 


All experiments were conducted locally using a MacBook Pro with M2 Pro processor and 32GB RAM. This hardware configuration represents typical development environments and demonstrates the feasibility of local LLM deployment for this type of automation.



\subsection{Artifact Validity}

\begin{figure}[htbp]
\centering
\includegraphics[width=1.0\linewidth]{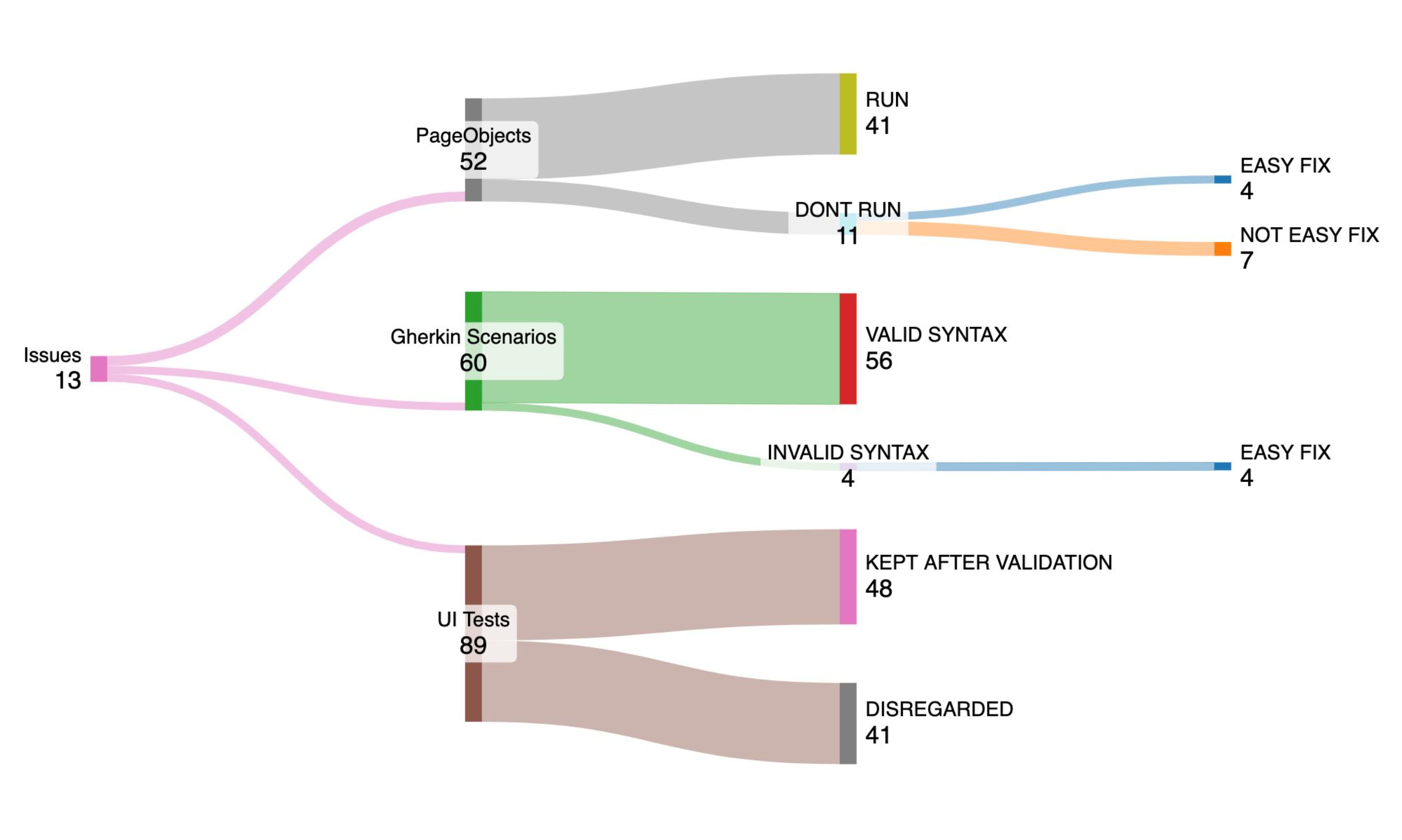}
\caption{Traceability Tree Showcasing the Different Artifacts}
\label{fig:artifact-qual-assess}
\end{figure}

Figure~\ref{fig:artifact-qual-assess} presents a traceability tree summarizing AToMIC’s ability to generate valid test artifacts across the full AT pipeline on 13 JIRA issues of the MyBMW project. 

Page Object generation proved to be the most challenging step. Of the 52 Page Objects generated, 78.8\% could be used without any manual modification. The remainder primarily suffered from technical issues, incorrect inheritance from the expected base class or imprecise function return types, which in several cases required moderate developer intervention. While many fixes were straightforward (such as changing a single function signature), 7 Page Objects required deeper structural changes.

Gherkin scenario generation achieved a 93.3\% success rate (56 out of 60 scenarios valid), as determined by a syntactic validation of each generated file. This verification step was carried out through manual checking, namely ensuring the presence of required Gherkin keywords (\texttt{Feature}, \texttt{Scenario}, and \texttt{Given/When/Then} blocks). The few invalid cases, concentrated within a single issue, typically involved missing background steps but remained easily readable and correctable within minutes.
In practice, Gherkin scenarios were used as documentation and validation artifacts, reviewed by stakeholders and developers to ensure alignment with acceptance criteria rather than executed directly as tests. This human-in-the-loop approach was essential, with participants confirming scenario clarity and relevance in the survey and artifact quality assessment (see Section~\ref{subsubsec:par-qua-ass}).

UI test script generation demonstrated robust outcomes after integrating the LLM-based validation step. Out of 89 initially generated test cases, 48 were retained after filtering by automatically cross-checking against the corresponding Gherkin scenarios. 
Every retained test was executed to confirm its syntactic correctness and practical executability within the test automation environment, with 100\% success. Subsequent manual analysis by the development team confirmed they were also semantically valid and thorough.

Overall, this traceability analysis confirms that AToMIC achieves high requirements coverage and practical validity across all major artifact types. The consistently high Gherkin and UI test success rates, in combination with human review and execution-based verification, provide strong evidence for the system’s applicability in industrial acceptance testing.

\subsection{Time Efficiency}

The efficiency gains achieved by AToMIC were substantial. Complete artifact generation, including Gherkin scenarios, Page Objects, and UI test scripts, took on average 259 seconds per issue when executed locally, as shown in Table~\ref{tab:efficiency}. 

\begin{table}[htbp]
\caption{Average execution time and token usage per issue.}
\label{tab:efficiency}
\centering
\small
\begin{tabular}{|l|c|c|c|}
\hline
\textbf{Artifact} & \textbf{Time (s)} & \textbf{Input Tokens} & \textbf{Output Tokens} \\
\hline
Page Objects & 142.6 & 21,911 & 1,979 \\
Gherkin Scenarios & 100.5 & 11,263 & 4,334 \\
UI Tests & 15.5 & 676 & 483 \\
\textbf{Total} & \textbf{258.6} & \textbf{33,850} & \textbf{6,796} \\
\hline
\end{tabular}
\end{table}

To better understand LLM usage, Table~\ref{tab:efficiency} also shows the number of input and output tokens processed by the language models.
As expected, Page Object generation accounted for the largest share of execution time and token consumption, followed by Gherkin scenario generation.   

It should be noted that execution times can be reduced by an estimated factor of 10x using cloud-based LLMs \cite{minions2025}, to approximately 26 seconds per issue, with an estimated cloud cost below \$0.01 per issue.

From a practical standpoint, developer feedback strongly supported the time savings observed. Several participants estimated that tasks such as Page Object creation, which previously consumed hours, could now be completed in minutes. Comments like “a full day saved” were not uncommon when discussing repetitive test maintenance workflows.


\subsection{Practitioner Survey and Qualitative Results}

To complement the technical evaluation, a practitioner survey was conducted using a convergent mixed-methods design. This approach combined quantitative ratings and qualitative feedback to assess the perceived quality, usability, and adoption potential of AToMIC-generated artifacts in a real-world industrial setting.

The survey was distributed to nine practitioners directly involved with the 13 JIRA issues. All participants reported being at least “somewhat familiar” with the selected issues, with 67\% indicating they were “very familiar.” The respondent pool was composed of seven developers, complemented by a Scrum Master and a Product Owner. In terms of experience, 44\% reported 1–3 years in their current role, 33\% had 4–5 years, and 22\% had over five years of experience. This distribution ensured that the feedback was grounded in both hands-on development practice and broader product and process perspectives.

\subsubsection{Artifact Quality}\label{subsubsec:par-qua-ass}

\begin{figure}[htbp]
\centering
\includegraphics[width=1.0\linewidth]{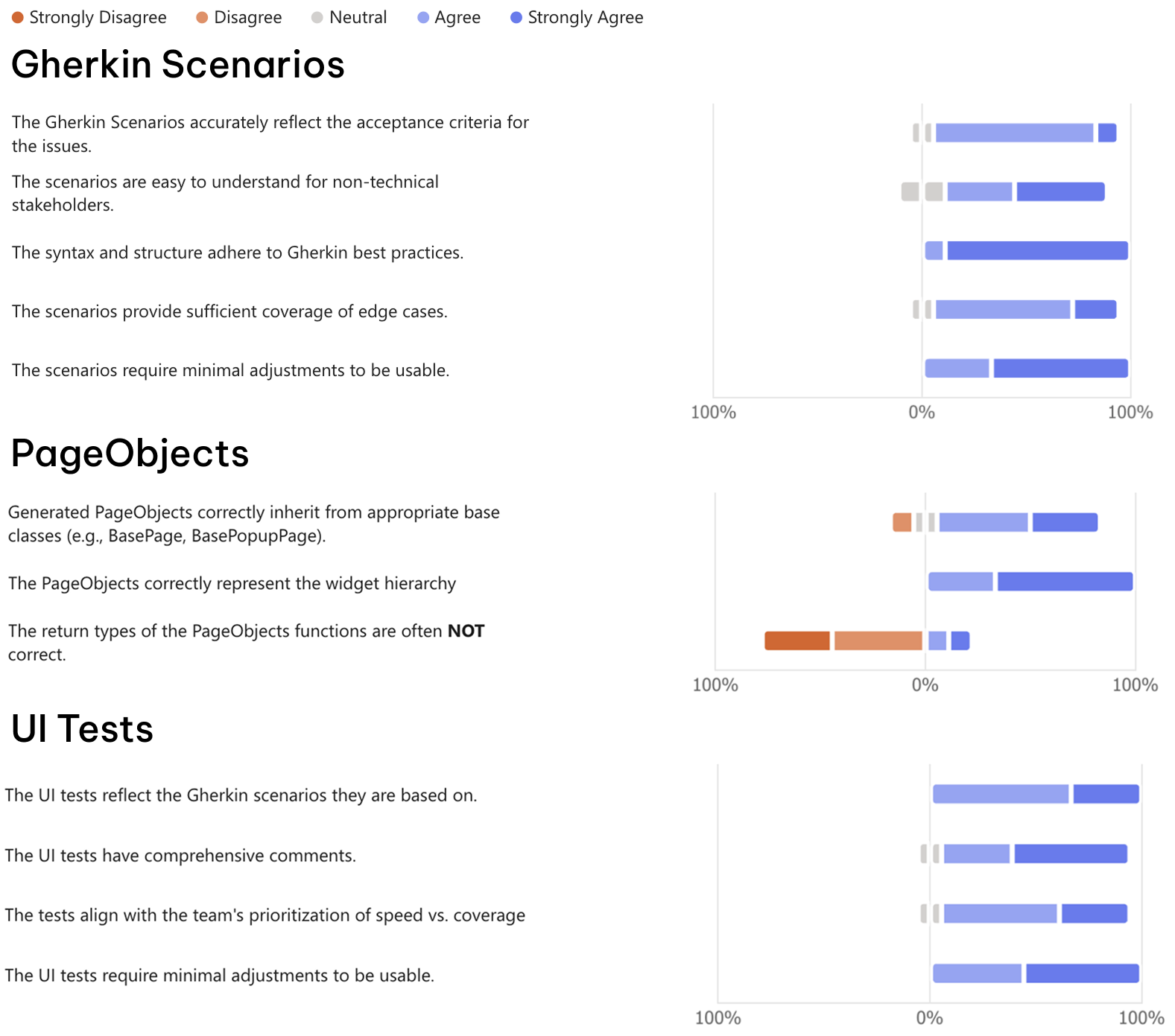}
\caption{Artifact Quality Assessment Survey Results}
\label{fig:quality-assess-val}
\end{figure}

The survey responses reflected strong confidence in the quality of AToMIC’s outputs across all artifact types, as shown in Figure~\ref{fig:quality-assess-val}.

Gherkin scenarios were universally well-received. All respondents (100\%) agreed that the generated scenarios accurately captured acceptance criteria, adhered to best practices, and required minimal adjustments. Additionally, 89\% found the scenarios accessible to non-technical stakeholders. While overall satisfaction was high, one participant noted that “huge projects with specific contexts bring some complexity to the LLM,” highlighting a potential limitation in highly specialized environments.

Page Objects also received generally positive feedback, though with some noted concerns. The main technical issues cited were related to inheritance hierarchy problems and return type mismatches. Despite these challenges, respondents consistently emphasized the time savings achieved. Several practitioners reported that AToMIC could save “up to one full day of work,” and one user remarked that “AToMIC will reduce a lot of workloads for the developers since the tests will be generated and only small adjusts will need to be done.”

UI test scripts received the most enthusiastic responses. All respondents (100\%) agreed that no manual adjustments were needed after generation (as confirmed in qualitative feedback), and that the generated test cases were both accurate and well-aligned with the intended scenarios.


\subsubsection{Usability and Adoption}

\begin{figure}[htbp]
\centering
\includegraphics[width=1.0\linewidth]{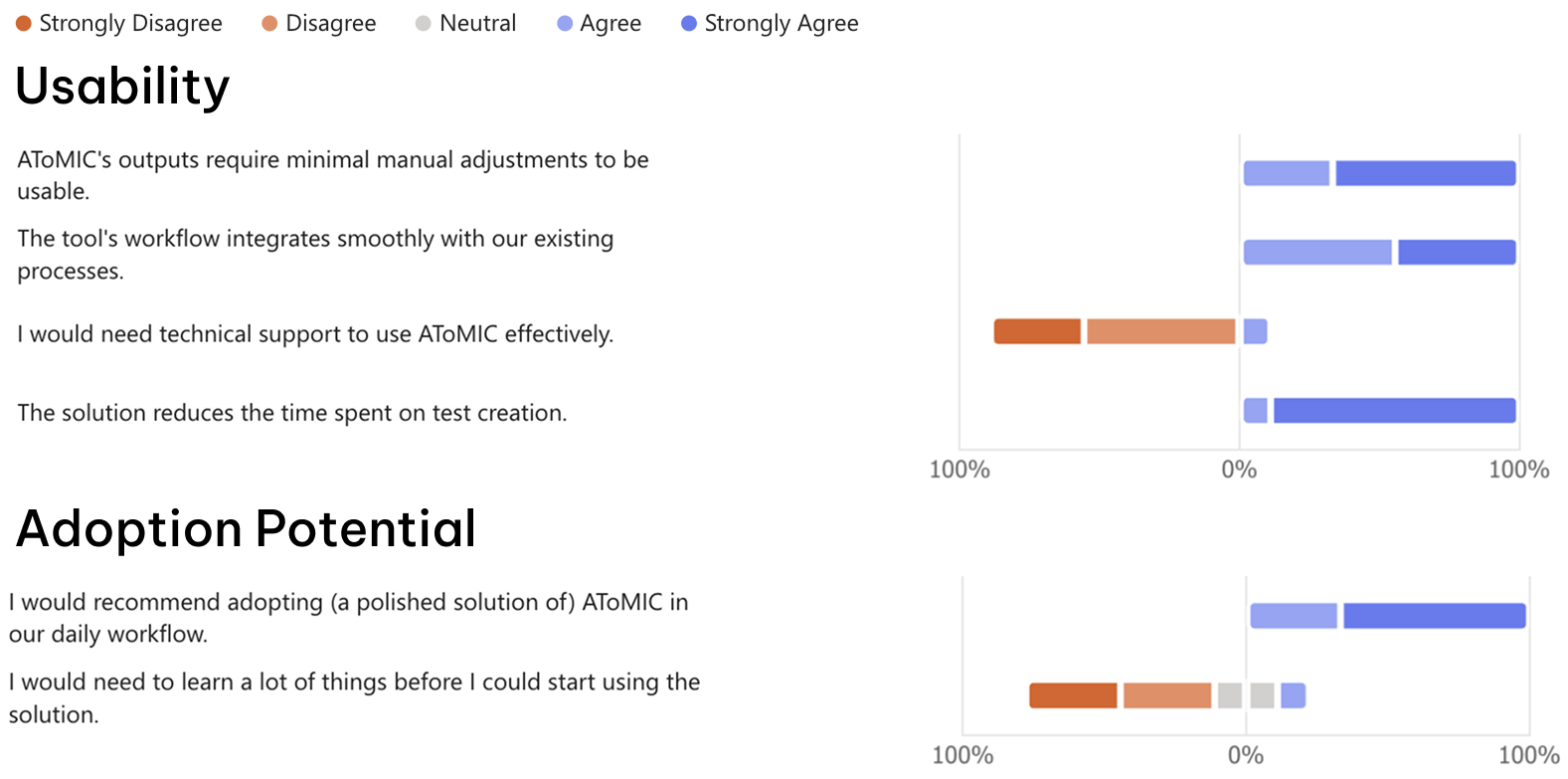}
\caption{Usability \& Adoption Potential Assessment Survey Results}
\label{fig:usal-adop-assess-val}
\end{figure}

Practitioners reported that AToMIC integrated smoothly into existing workflows. All participants agreed that the tool reduced the time required for test creation, and only 11\% felt that technical support was needed to use it effectively (see Figure~\ref{fig:usal-adop-assess-val}). The learning curve was perceived as low, with most respondents disagreeing that significant training would be necessary.

Importantly, every participant indicated they would recommend incorporating AToMIC into daily development practice, highlighting its immediate applicability and value within the team’s tooling ecosystem. 

\subsubsection{Qualitative Insights}

Open-ended responses were analyzed using a structured thematic approach, revealing three core themes.

\textbf{Time and Productivity Gains.} The most frequently mentioned benefit was the time saved by automating repetitive tasks. Developers cited examples such as “saving about one full day of work” on Page Object generation. AToMIC was described as freeing practitioners from “boring” manual tasks, allowing them to focus on more meaningful work.

\textbf{Quality and Reusability of Outputs.} Several respondents highlighted that AToMIC provides a reliable starting point for Gherkin scenarios, Page Objects, and UI tests. One participant noted that it “helps in reducing the time needed for development” by offering a foundation that require minimal polishing.

\textbf{Requests for Workflow Enhancements.} While feedback was largely positive, respondents suggested improvements such as tighter integration with test management tools like Xray and JIRA, enhanced automatic file placement, and better prompt contextualization through expanded use of story descriptions. These requests point to natural next steps for improving usability and workflow alignment.

Finally, feedback revealed a balanced understanding of AToMIC’s role in the development lifecycle. Practitioners consistently framed it as an intelligent assistant rather than a replacement for human judgment. Several noted that while the system performs well, it still struggles with complex, domain-specific contexts, particularly in large projects. Nonetheless, they saw clear value in the system’s current capabilities and expressed confidence in its future potential.

\subsection{Answers to Research Questions}

This subsection synthesizes the quantitative and qualitative findings to answer the research questions guiding AToMIC's empirical evaluation.

    \textbf{RQ1: To what extent are the generated artifacts accurate and complete when using LLMs for automating AT in mobile apps?}


Empirical results show that AToMIC generates highly accurate and complete artifacts for industrial-scale Flutter apps. Gherkin scenarios were 93.3\% syntactically correct, and all artifacts were usable by both technical and non-technical stakeholders. Page Objects and UI tests reached executability rates of 78.8\% and 100\% (after LLM filtering), with minimal manual fixes. Practitioners confirmed strong traceability and semantic alignment with requirements.

    \textbf{RQ2: Is the process of generating acceptance test scripts using LLMs more time-efficient compared to traditional manual methods?}


Quantitative results show that AToMIC significantly reduces test generation time—averaging 259 seconds per issue locally. Estimated cloud execution times suggest a potential 10× speed-up, with negligible cost. In contrast, practitioners estimated a full day of manual effort per issue, yielding over 95\% time savings. These gains match or exceed results from comparable LLM-based studies~\cite{10.1145/3695988}, with no major bottlenecks and smooth workflow integration reported. It is important to note that while automation inherently provides time-efficiency benefits, these gains would not be as meaningful without the high quality of the generated artifacts.

    \textbf{RQ3: How accessible and understandable are the generated acceptance test scripts for individuals with varying levels of technical expertise?}

Practitioner survey results show that AToMIC’s artifacts are accessible to both technical and non-technical users. Gherkin scenarios were praised for their clarity and traceability, while Page Objects and test scripts were well-structured and needed only minor adjustments. All participants agreed that the generated artifacts ease test comprehension and maintenance.

 \textbf{RQ4: What are the main challenges and limitations encountered when applying LLMs to acceptance test generation for mobile applications, and how do they impact the overall process?}

    The evaluation revealed several challenges, reflecting both model and workflow limitations. Page Object generation sometimes required minor manual fixes, such as correcting base class inheritance and method signatures.  The complexity and modularity of large-scale Flutter projects complicated navigation mapping, which required strict standardization and developer discipline in widget key naming. Privacy constraints also limited LLM execution to local environments, preventing the use of the latest large-scale commercial models. Additionally, some users suggested improvements like integration with test management tools. Despite these issues, practitioners consistently found that AToMIC’s productivity gains, artifact quality, and strong traceability outweighed the drawbacks.

In summary, the evaluation affirmatively answered all research questions: AToMIC enables efficient, high-quality, and accessible generation of acceptance test artifacts for industrial Flutter projects, while addressing workflow, technical, and organizational challenges observed in real-world deployments.

\subsection{Threats to Validity}




\textbf{Internal Validity.}  
To ensure consistency, all experiments used the same hardware, software stack, and local LLM deployment. However, minor manual steps—such as Page Object mapping setup and artifact corrections—may have introduced some subjectivity. While human interventions were minimised, their exact frequency was not systematically tracked, which could slightly overstate productivity gains. Transparent reporting, shared artifacts, and independent practitioner surveys helped mitigate bias and enhance reliability.



\textbf{External Validity.}  
While our evaluation focuses on Flutter applications within BMW's development environment, AToMIC's architectural design supports broader applicability. More than half of the workflow components are platform-agnostic. Issue retrieval, commit analysis, navigation modeling, and Gherkin generation can all be applied to any mobile framework. The primary adaptation challenge lies in PageObject generation, which requires framework-specific UI abstraction patterns. Although the MyBMW app is representative of complex, modular systems, broader evaluations across diverse projects and organizational contexts would strengthen the generalizability of our findings. Future work will assess AToMIC's effectiveness on other platforms and develop framework-specific PageObject generation modules to enable wider adoption.



\textbf{Construct Validity.}  
Artifact quality and efficiency were assessed through syntactic checks and practitioner feedback.
Participants’ prior familiarity with the workflow may have positively influenced perceptions. To reduce bias, surveys were administered independently and supported by objective metrics.






In conclusion, while the study employed multiple strategies to mitigate threats—including consistent setup, detailed documentation, and independent evaluation—some limitations remain. Future research will expand validation scope and further reduce manual intervention.

\section{Conclusions and Future Work}

This work addressed the persistent challenge of manual, time-consuming, and error-prone acceptance test generation in mobile software development, particularly for cross-platform frameworks like Flutter. We presented AToMIC, a modular system that automates the generation of Gherkin scenarios, Page Objects, and UI test scripts directly from requirements and code changes, leveraging locally deployed LLMs. AToMIC integrates into industrial CI/CD pipelines and supports realistic mobile workflows through systematic Page Object abstraction and navigation modelling.

The empirical evaluation demonstrates that AToMIC achieves high efficiency and quality in automated acceptance artifact generation for complex industrial Flutter applications. Quantitative results showed significant time savings—over 95\% compared to manual test creation—and high artifact correctness and usability. Practitioner feedback confirmed the value of the generated artifacts in terms of productivity, clarity, traceability, and ease of adoption. These findings affirm the feasibility and impact of LLM-driven test automation in real-world settings.

Our main contributions include: (i) an end-to-end automated workflow from issue tracking and version control to executable test artifacts, (ii) formalized methods for Page Object generation and navigation modeling, (iii) prompt engineering and LLM deployment strategies tailored for privacy-constrained environments, and (iv) a validated implementation based on thirteen real-world issues in the MyBMW app.

While AToMIC performs effectively in production, several directions remain for future work. Enhancing Page Object generation to better handle complex widget structures, return type inference, and dynamic flows is a key area. Adapting the system to newer, high-capacity LLMs—cloud-based or on-premises—may further boost quality and scalability, provided compliance with data privacy constraints. Broadening applicability to other platforms beyond Flutter (e.g., native Android/iOS or other cross-platform frameworks) would expand its impact. 


Overall, AToMIC demonstrates that LLM-based automation can significantly streamline the creation of high-quality, maintainable AT artifacts, reducing manual effort while supporting collaboration across technical and non-technical stakeholders in industrial software engineering.

\section*{Acknowledgments}

This work was financially supported by: UID/00027 of the LIACC  - Artificial Intelligence and Computer Science Laboratory - funded by Fundação para a Ciência e a Tecnologia, I.P./ MCTES through the national funds.

\appendix[Artifacts Generation Example]\label{app:example}

This appendix illustrates AToMIC’s generation workflow for a JIRA issue related to the charging adapter functionality in the MyBMW app. For confidentiality reasons, only some excerpts of the artifacts are shown.  

The Jira issue has the following characteristics:

\section{JIRA Issue Overview}
\begin{itemize}
    \item \textbf{Key}: NWAP-165701
    \item \textbf{Summary}: Add link to Driver's Guide
    \item \textbf{Labels}: Backend, Mobile
    \item \textbf{Acceptance Criteria}:
    \begin{enumerate}
        \item The "BMW Driver's Guide" link on the "About Adapters" page redirects to the BMW Driver's Guide app if installed.
        \item If the app is not installed, the link redirects to the App Store.
        \item When redirected, the user should land on the Charging Vehicle section in the correct language.
    \end{enumerate}
    \item \textbf{Description}:
    \begin{lstlisting}[breaklines=true]
Goal 
Create deeplink to Driver's Guide.

Background 
In one of our onboarding screens, we provide a link to the Driver's Guide to inform users about NACS adapters. We need to create a deeplink for this purpose.
[...]
    \end{lstlisting}
\end{itemize}

\subsection{GitHub Commit Identification \& File Filtering}

AToMIC identified a single commit directly associated with this issue. 
The commit introduced modifications exclusively in the 'adapters\_configuration' module, affecting 4 .dart files.


\subsection{Page Object Generation}

From the code changes, AToMIC generated 3 Kotlin Page Objects for the updated screens. 
Listing~\ref{lst:atomic-add-adapter-page} shows an extract of one of such pages - the `AddAdapterPage`, encapsulating the UI elements and user interactions involved in adapter selection and saving.


\begin{lstlisting}[breaklines=true,language=Java,caption={Generated Page Object for the Add Adapter Page},label={lst:atomic-add-adapter-page}]
package pages.chargingequipments.adaptersconfiguration
import pages.base.BaseTabPage
...
class AddAdapterPage<T : BaseTabPage>(previousPage: AdaptersMainPage<T>) : BasePopupPage<AdaptersMainPage<T>>(previousPage) {
    ...
    override fun ensurePageVisible() {
        driver.wait(WaitingConstants.PAGE_SWITCH_WAITING_TIME, selectAdapterListItemSelector)
    }
    fun selectAdapter(): AboutAdaptersPage<T> {
        driver.scrollAndClick(selectAdapterListItemSelector)
        return AboutAdaptersPage(this)
    }
    fun save(): AdaptersMainPage<T> {
        driver.wait(WaitingConstants.PAGE_SWITCH_WAITING_TIME, selectAdapterListItemSelector)
        driver.waitAndClick(saveButtonSelector)
        previousPage.ensurePageVisible()
        return previousPage
    }
}
\end{lstlisting}

\subsection{Navigation Path Discovery}

AToMIC identified several relevant navigation paths required to reach the target feature. Listing~\ref{lst:atomic-discovered-paths} outlines a subset of these discovered paths, each specifying page transitions and actions.


\begin{lstlisting}[breaklines=true,language=bash,caption={Discovered Navigation Paths for NWAP-165701},label={lst:atomic-discovered-paths}]
Path 1:
  VehicleTabPage 
  -> ElectricMobilityPage (via charging)
  -> AdaptersMainPage (via adaptersConfiguration)
  -> DriversGuide (via openDriversGuide)

Path 2:
  VehicleTabPage 
  -> MyVehicleListPage (via enterGarage)
  -> ManualVinEntryInfoPage (via enterToManualVinEntryInfoPage)
  -> ManualVinEntryPage (via enterToAddVehicleEnterVinPage)
  -> BmwIntroPage (via submitVin)
  -> VehicleTabPage (via submitLetsGoBtn)
  -> ElectricMobilityPage (via charging)
  -> AdaptersMainPage (via adaptersConfiguration)
  -> DriversGuide (via openDriversGuide)
[...]
\end{lstlisting}

\subsection{Gherkin Feature File Generation}

The Gherkin generator produced a feature file with two user flows: with and without the BMW Driver’s Guide app installed, as shown in Listing~\ref{lst:gherkin-drivers-guide}, expressing the acceptance criteria as executable test scenarios.

\begin{lstlisting}[breaklines=true,language=Gherkin, caption={Generated Gherkin Feature File for NWAP-165701},label={lst:gherkin-drivers-guide}]
Feature: Add link to BMW Driver's Guide

Scenario: User has BMW Driver's Guide installed
  Given The user clicks on the "Add Adapter" button in the About Adapters widget
  When The BMW driver's guide link appears in the About Adapters widget
  Then The user is redirected to the BMW Driver's Guide app with correct language displayed

Scenario: User does not have BMW Driver's Guide installed
  Given The user clicks on the "Add Adapter" button in the About Adapters widget
  When The BMW driver's guide link appears in the About Adapters widget
  Then The user is redirected to the App Store with correct language displayed
\end{lstlisting}

\subsection{Generated UI Test Class}

Finally, AToMIC compiles the generated Page Objects and navigation paths into an executable UI test class. Listing~\ref{lst:atomic-uitest-class} presents an excerpt of a generated test case, following Path 1 in Listing~\ref{lst:atomic-discovered-paths}.

\begin{lstlisting}[breaklines=true,language=Java,caption={Generated UI Test Class for NWAP-165701},label={lst:atomic-uitest-class}]
package uitest.vehicletab

import pages.vehicle.VehicleTabPage
...

@Priority1
class AddLinktoDriversGuideinAboutAdaptersPage : BaseUiTest() {

  private lateinit var vehicleTabPage: VehicleTabPage

  public override fun setup() {
    vehicleTabPage = initApp(). ...
  }

  // Scenario: When all adapters are configured and the user navigates via charging link to About Adapters page
  @RetryingTest(2)
  fun driversGuideTest() {
    vehicleTabPage
      .charging().adaptersConfiguration()  // Navigate from Charging to Adapters Configuration
      .openDriversGuide() // Open the Driver's guide
      .back()  // Go back to the previous page (Charging)
      .back()  // Go back again to reach the initial state (Vehicle Tab Page)
  }
}
\end{lstlisting}

\bibliographystyle{IEEEtran}  
\bibliography{myrefs}

\end{document}